\documentstyle[12pt,aaspp4,epsfig]{article}
\def\deg{\ifmmode^\circ\else$^\circ$\fi}

\def\mic{~$\mu$m}

\def\h0{H$_0$}
\def\q0{q$_0$}

\def\arcs{\ifmmode {''}\else $''$\fi}
\def\arcm{\ifmmode {'}\else $'$\fi}
\def\parcs{\sa=.07em \sb=.03em
     \ifmmode $\rlap{.}$^{\scriptscriptstyle\prime\kern -\sb\prime}$\kern -\sa$
     \else \rlap{.}$^{\scriptscriptstyle\prime\kern -\sb\prime}$\kern -\sa\fi}
\def\parcm{\sa=.08em \sb=.03em
     \ifmmode $\rlap{.}\kern\sa$^{\scriptscriptstyle\prime}$\kern-\sb$
     \else \rlap{.}\kern\sa$^{\scriptscriptstyle\prime}$\kern-\sb\fi}

\def\Msun{M$_{\odot}$}
\def\Myr{\Msun/yr}
\def\kp{\mbox{{\rm K}$^{\prime}$}}

\def\han {\mbox{{\rm H}$\alpha$}}
\def\hb {\mbox{{\rm H}$\beta$}}

\def\oiii{[O~{\sc iii}]}
\def\oiiifive{[O~{\sc iii}]~$\lambda 5007$}
\def\oii{[O~{\sc ii}]}

\def\ha{\han}

\def\brg {{\rm Br}$\gamma$}

\def\spose#1{\hbox to 0pt{#1\hss}}
\def\simlt{\mathrel{\spose{\lower 3pt\hbox{$\mathchar''218$}}
     \raise 2.0pt\hbox{$\mathchar''13C$}}}
\def\simgt{\mathrel{\spose{\lower 3pt\hbox{$\mathchar''218$}}
     \raise 2.0pt\hbox{$\mathchar''13E$}}}
\def\lsim{\rlap{$<$}{\lower 1.0ex\hbox{$\sim$}}}
\def\gsim{\rlap{$>$}{\lower 1.0ex\hbox{$\sim$}}}

\begin{document}


\title{Measurement of \oiii~Emission in Lyman Break Galaxies
\footnote{
Data presented herein were obtained at the W.M. Keck Observatory, 
which is operated
as a scientific partnership among the California Institute of Technology, 
the University of California and the National Aeronautics and Space 
Administration.  The Observatory was made possible by the
generous financial support of the W.M. Keck Foundation.
}}

\author{
Harry I. Teplitz\altaffilmark{2,}\altaffilmark{3},\\
Matthew A. Malkan\altaffilmark{4}, Charles C. Steidel\altaffilmark{5}
Ian S. McLean\altaffilmark{4}, 
E. E. Becklin\altaffilmark{4}, 
Donald F. Figer\altaffilmark{6},
Andrea M. Gilbert\altaffilmark{7}
James R. Graham\altaffilmark{7},
James E. Larkin\altaffilmark{4},
N. A. Levenson\altaffilmark{8}, 
Mavourneen K. Wilcox\altaffilmark{4}}
\altaffiltext{2}{Laboratory for Astronomy and Solar Physics, Code 681, Goddard
Space Flight Center, Greenbelt MD 20771}
\altaffiltext{3}{NOAO Research Associate}
\altaffiltext{4}{Department of Physics and Astronomy, 
University of California, 
                 Los Angeles, CA, 90095-1562 }
\altaffiltext{5}{
California Institute of Technology, Palomar Observatories,
Pasadena,  CA, 91125}
\altaffiltext{6}{Space Telescope Science Institute, 
                  3700 San Martin Dr., Baltimore, MD 21218 }
\altaffiltext{7}{
Department of Astronomy,  
University of California, Berkeley,
601 Campbell Hall,
                 Berkeley, CA, 94720-3411}
\altaffiltext{8}{Department of Physics and Astronomy,  
Johns Hopkins University,
                 Baltimore, MD  21218}

\begin{abstract}
  
  Measurements of \oiii~emission in Lyman Break galaxies (LBGs) at $z
  > 3$ are presented.  Four galaxies were observed with narrow-band
  filters using the Near-IR Camera on the Keck I 10-m telescope.  A
  fifth galaxy was observed spectroscopically during the commissioning
  of NIRSPEC, the new infrared spectrometer on Keck II.  The
  emission-line spectrum is used to place limits on the metallicity.
  Comparing these new measurements with others available from the
  literature, we find that strong oxygen emission in LBGs may suggest
  sub-solar metallicity for these objects.  The \oiiifive~line is
  also used to estimate the star formation rate (SFR) of the LBGs.
  The inferred SFRs are higher than those estimated from the UV
  continuum, and may be evidence for dust extinction.  

\end{abstract}


\section{Introduction}

Recent advances in infrared instrumentation have extended the study of
high redshift ($z>2$) galaxies to rest-frame optical wavelengths.  The
Lyman Break Galaxies (hereafter LBGs; c.f. Steidel et al. 1996) have
been well characterized from their rest-frame far-UV spectra, using
optical spectrographs.  With the commissioning of NIRSPEC on Keck II
and ISAAC on VLT, we have begun to obtain rest-frame optical spectra.
The far-UV spectra of LBGs are strikingly similar to nearby
starbursts; for example Steidel et al. (1996) compare directly with
the Wolf-Rayet galaxy NGC 4214 (Leitherer et al. 1996).  The
rest-frame optical spectra of LBGs also resemble the spectra of low
redshift irregulars and starbursts.  The familiar bright emission-line
diagnostics are present in LBG spectra (\ha, \oiiifive, \hb,
\oii~$\lambda 3727$).  The first rest-frame optical spectra of LBGs
were obtained by Pettini et al. (1998; hereafter P98).  A spectrum of
the gravitationally lensed LBG MS1512-cB58 (see Yee et al. 1996) was
later obtained by Teplitz et al. (2000; hereafter T2000).  At $z>2.8$,
\ha~redshifts out of the K-band, and so the most easily observed emission line in
most LBGs is \oiiifive, which is a strong line in star forming
galaxies.

In this paper we present Keck measurements of \oiii~in a sample of
five typical LBGs.  Four of the objects were imaged in narrow band
filters centered on the expected wavelength of \oiiifive~using the
Near-Infrared Camera (NIRC; Matthews \& Soifer 1994) on the Keck I
10-m telescope.  A fifth LBG was observed from Keck II with the new
near-IR spectrograph (NIRSPEC), and a K-band spectrum was obtained
(rest-frame 4600-5200\AA).

In section 2  the observations are presented and the data reduction techniques 
are explained.  Section 3 outlines the observational results, and section 4 
presents a discussion of their implication for understanding LBGs.  The conclusions
are summarized in section 5.  Unless 
otherwise noted, a cosmology of (\h0=75 km s$^{-1}$~Mpc$^{-1}$,\q0$=0.1$,$\Lambda=0$) 
is assumed.

\section{Observations}

\subsection{Imaging}
\label{sec:  imaging}

Targets were chosen from the database of spectroscopically confirmed
LBGs observed by Steidel et al.  (see e.g. Steidel et al. 1996).  We
will use their nomenclature to refer to individual objects.  Table
\ref{tab: observations}~lists the imaging targets.  Targets located in
the so called ``Groth-Westphal strip'' (Groth et al. 1994), will be
referred to as ``WESTPHAL-CC\#'' or simply WCC-\#. The other two
fields from which LBGs were selected were centered on the radio
galaxies 3C324 and B20902.  Narrow-band fields were chosen to center
on LBGs with redshifts that place the redshifted \oiiifive~line within
0.5\% of 2.16\mic, which is the central wavelength of the \brg~filter.

Imaging observations were taken with the Near IR Camera (NIRC;
Matthews \& Soifer 1994) on the Keck I telescope, on the nights of
April 7-8, 1999.  Broad-band \kp~images were taken to measure the
continuum flux in the objects, and narrow-band images were taken to
measure the redshifted \oiiifive~flux.  The narrow-band filter used
for these observations was the standard \brg~filter, centered at
2.16\mic~with a width $\Delta \lambda / \lambda = 0.01$.  Teplitz,
Malkan, \& McLean (1998, hereafter TMM98) confirmed the width of the
filter from standard star observations.  Table \ref{tab:
  observations}~lists the integration times for each observation. As
shown in the table, the WESTPHAL-CC18 and 3C324 fields were observed
to greater depth.  Both nights were photometric.  In each case the
broad- and narrow-band observations were taken as close in time as
possible.  Typically integrations were taken for 27 minutes on the broad-band, then 90
minutes narrow-band.  The deeper fields were observed on both nights,
but in those cases both broad- and narrow-band observations were made
each night.  Broad band observations were taken in 60 second
exposures consisting of four coadds of 15 seconds each.
Narrow-band observations were taken in single 240 second exposures.
The seeing varied during the nights from $\sim 0.6$\arcs~to $\sim
0.8$\arcs~(with the worst seeing at the end of the second night).
UKIRT faint standard stars (Casali \& Hawarden 1992) were observed
periodically during the nights for photometric calibration of the
broad-band images.  It is difficult to calibrate narrow-band
photometry onto flux units, so we measure the ratios of detected
counts in the broad- and narrow-bands and then convert to narrow-band
flux excess from knowledge of the width and throughput of the filter
(see TMM98).  Bunker et al. (1995) adopted the same approach but have
calibrated the narrow-band magnitude with the broad-band zeropoint so
that featureless continuum objects have a broad-minus-narrow-band
color that is identically zero.

The images were reduced with the same procedure described in TMM98,
which will only be summarized here.  We obtained images in a sequence
of ``dithered'' exposures, offsetting the telescope between exposures
in a 3$\times$3 grid spaced by a few arcseconds.  The data were
reduced by dividing a twilight flat into each image and then
subtracting from it a running median sky frame created from the nine
other exposures taken closest in time.  Objects were identified using
the SExtractor (Bertin \& Arnouts 1996) software.

Photometry was performed in circular
apertures of 10 pixels (1.5\arcs) diameter (approximately 2.5 times
the seeing disk), which corresponds to 11 kpc at z=3.3. The size of
the aperture was chosen primarily to obtain optimal signal to noise
(see Thompson et al. 1995 and Howell 1989).  This aperture should
encompass the light from a typical LBG (Lowenthal et al. 1997).  The
same aperture was applied to broad and narrow-band exposures. 

Photometric errors were estimated from aperture photometry performed
on random positions in the frame. The signal to noise ratio (SNR) for
the individual LBGs in both bands is given in Table \ref{tab: nb_lbg}.
The confidence intervals in the narrow-minus-broad band color were
estimated from Monte Carlo simulations (see TMM98).  These simulations
generated narrow- and broad-band magnitudes for line-free objects
having the Gaussian errors measured in the real data.  

Emission line objects are identified as lying above the
$3\sigma$~confidence interval in the narrow-minus-broad band counts.
The quantity of interest, specifically, is $\Delta m$, the excess flux
in the narrow-band in magnitudes.  Featureless continuum objects will
have the same ratio of detected counts between the broad- and
narrow-bands.  This ratio depends simply on the width of the filter in
wavelength and on its efficiency in transmitting light.  In TMM98 the
\brg~filter in NIRC was found transmit $\sim 1/24$~of the light
compared to the \kp~filter.  Thus in measured (raw) counts, a
featureless object will have $m_K - m_{n.b.} \sim -3.45$.  An emission
line will be seen as excess flux in the narrow band, and we can
calculate the narrow-band excess, $\Delta m$, which is the difference
between the raw broad-minus-narrow-band color of the object compared
to the color of a featureless continuum:
\begin{equation}
\Delta m = m_{K} - m_{n.b.} + 3.45
\end{equation}

Since the narrow-band filter is centered at 2.16\mic, with a width 
of 1\%~in wavelength, our imaging observations measure the redshifted
\oiiifive~line.  These observations should be uncontaminated by 
the \oiii~$\lambda 4959$~line, which is 0.96\%~away in wavelength.  
We must convert from the observed $\Delta m$~to line flux.  Using the standard procedure
for narrow-band emission line measurements, we first calculate the equivalent
width (EW) following, for example, Bunker et al. (1995).  We
find that 
\begin{equation}
EW_{obs} \simeq \Delta \lambda _{n.b.} \left( 10^{0.4\Delta m} - 1 \right)
\end{equation}
where $\Delta \lambda _{n.b.}$~is the width of the narrow-band filter
in \AA, and assuming that the emission line is a negligible contribution
to the broad-band flux.  The EW is multiplied by the observed
broad-band flux.  Thus, errors on the line flux stem from the
photometric errors, and the error propagation is handled in the usual
manner.

\subsection{Spectroscopy}

The spectroscopic target (WESTPHAL-CC13; $z\simeq 3.406$) was chosen for its redshift
which places \oiii~and \hb~between the strong atmospheric OH emission
lines, and for its brightness in the rest-frame UV.

The Near IR spectrum of Westphal-CC13 was obtained using the NIRSPEC
instrument on the 10-m W. M. Keck II telescope.  The delivered
spectrograph is described in detail in McLean et al. (1998, 2000).  It
is a cross-dispersed echelle spectrograph, with a $1024\times
1024$~pixel (ALADDIN2) InSb detector.  A flat mirror in place of the
Echelle grating allows lower resolution ($R\sim 2000$) spectra to be
taken. In this mode, each detector pixel corresponds to 0.144\arcs~in
the spatial direction and 0.19\arcs~in the spectral direction.  A
$256\times 256$~pixel (PICNIC) HgCdTe array provides simultaneous
slit-viewing when filters in the $1-2.5$\mic~bands are used.  The slit
viewing camera (SCam) has a plate scale of 0.18\arcs/pixel.

The spectrum was obtained in the low resolution, long-slit mode
through the 42\arcs~long slit.  A slit width corresponding to 0.57\arcs~(three
pixels) at the InSb detector was chosen, yielding a final resolution of
R$\sim 2000$.  The spectrum was obtained in three 900 seconds
integrations, separated by small ($\sim 8$\arcs) nods along the slit.
A mechanical problem with a filter wheel reduced the throughput of the
observations on this run, such that the 90 minutes of integration was
equivalent to 45 minutes with NIRSPEC's usual peak performance.  The
seeing was very good ($\sim 0.35$\arcs), so most of the light from the
object should have been transmitted by the slit.  The object was
acquired using its known position with respect to a nearby bright
star.  The star was centered in the slit, with the instrument's
internal image rotator set to the position angle that allowed the star
and WESTPHAL-CC13 to be observed simultaneously (243\deg).  

We reduced the data using custom software written for the NIRSPEC
instrument in the IDL language.  First, a halogen lamp flat field
image was used to remove pixel-to-pixel variations in the detector
response.  Known bad pixels and obvious cosmic rays are identified and
fixed using an IDL version of standard IRAF\footnote[6]{IRAF is
  distributed by NOAO, which is operated by AURA Inc., under contract
  to the NSF} routines for this purpose.  The next step was to
subtract a sky image, constructed from the nodded object frames.  Due
to the prevalence of OH lines in the spectrum, which change
independently from each other with time, the sky frame had to be
scaled to the sky level at the time of the individual observations.
Finally, the two dimensional spectral image was rectified onto a
linear space vs. wavelength grid.

Rectification of the 2D spectrum was the most complicated part of
the data reduction process.  Raw spectral data are rotated with
respect to detector pixels, and distorted by many pixels in
wavelength near the edges of the frame.  The wavelength scale for the
final grid was determined from measurement of Neon and Argon arc lamp
spectra taken immediately following the observations.  The spatial
distortion was calculated from measurement of the pixel spacing
between nodded spectra of a standard star. Spatial distortion was
corrected first by linearly interpolating each row of the object frame.
Next, each column was independently interpolated onto the wavelength
scale using a 3$^{rd}$-order fit to the arc lines.  The final
rectification provided wavelength calibration good to 0.99\AA.  The
output pixel scale was 4.16\AA/pixel.  
Unresolved Argon arc-lamp lines had FWHM$=11.3$\AA, for a final
resolution of, for example, R$= 1940$~at 2.2\mic. 

After rectification the frames were registered to coadd the nodded data.  The
nod distance in the new spatial coordinates was recovered by fitting a
Gaussian to the spectrum of the bright reference star.  The final 2D
spectrum was a weighted mean of the registered images, with weights
calculated from the integral under the Gaussian fits to the emission
line.  Some regions of the spectrum remain unrecoverable due to the
night sky lines.  
We extracted 1D spectra using the stellar continuum shape to define an optimal
extraction vector.  This vector was used to weight each (spatial) row of the
spectrum, and then the object region was summed.  

Extracted spectra were divided by the atmospheric absorption spectrum.
We obtained this spectrum from observation of a mostly featureless star
(SAO 46659, an A0V star).  The stellar data were reduced in the same manner as
the spectra of WESTPHAL-CC13, divided by a Kurucz (1993) model atmosphere. 
The K magnitude of the star was extrapolated from 
optical photometry based on its stellar type.  The sensitivity of the
calibrated spectrum was then obtained as a function of flux density
per data number per pixel per second and this calibration was applied
to the spectrum of WESTPHAL-CC13.  It is not possible to estimate the equivalent
widths of the lines due to the lack of continuum signal detection.  However,
P98 find that the equivalent widths of \oiiifive~and \hb~are large ($> 40$\AA~in 
the rest frame).  

\section{Results}

Figure \ref{fig: cm}~plots the narrow-band excess, $\Delta m$~(see
section 2.1), against the \kp~magnitude for all the objects in the
fields of the LBGs. The known LBGs are circled in the figure. A strong 
narrow-band excess revealed by plotting this difference identifies an
emission line object (see for example Thompson, Mannucci, \& Beckwith
1996).  The identification of an emission line object depends on the
$3\sigma$~limits propagated from the photometric errors (section
\ref{sec: imaging}); an object above the limit is an emission line
detection.  The $3\sigma$~limits are plotted separately for the two
deeper fields (3C324 and WESTPHAL-CC18) and the shallower ones (B20902
and WESTPHAL-CC8).  The figure shows that no new emission-line sources
are detected.  The four known LBG sources are faint at the depth of
this survey, and are near the detection limits.  However, two of them
appear to show significant ($>3\sigma$) \oiii~emission as expected for
young, sub-solar metallicity starforming galaxies (c.f. Kobulnicky et
al. 1999; hereafter KKP99).  The brightest of the LBGs shows little
\oiii~emission compared to the others.  These narrow-band excesses are
given in Table \ref{tab: nb_lbg}.

Figure \ref{fig: kspec}~shows the K-band spectrum of WESTPHAL-CC13.
The \oiiifive~and \oiii~$\lambda 4959$~lines are detected, but
\hb~is not, though an upper limit has been measured from the
data.  Table \ref{tab: wcc13}~lists the measured properties of
WESTPHAL-CC13.  Table \ref{tab: all_lbg}~compares the measured and
inferred properties of the narrow-band imaging and spectroscopic
sample, together with other LBGs available in the literature.  Note
that in table \ref{tab: all_lbg}~there is a second object (B20902-C6)
near one of our narrow-band LBG targets (B20902-M11), but that it lies
outside the redshift range covered by the narrow band filter; its flux 
was measured spectroscopically by P98.

\section{Discussion}

We assume that the source of the ionization for the oxygen lines we measure are 
the stars in the LBGs rather than the presence of active galactic nuclei (AGNs).
The rest-frame UV spectra of the LBGs show no evidence for AGN activity, such 
as the high ionization lines that might be expected.  The 
possibility that a dust enshrouded AGN could be present is unlikely given the
relatively blue continuum colors of the galaxies.  Future rest-frame optical 
spectroscopy will rule out the possibility if other non-stellar lines, such as
[Ne~{\sc V}]3425\AA, are not seen, or based on the line ratios of bright
optical emission lines (see Rola, Terlevich \& Terlevich 1997, who separate starbursts
from Seyferts based on the ratios of \oii $\lambda$3727, [Ne {\sc III}]$\lambda$3869,
\hb, and \oiiifive).  

\subsection{The metallicity of WESTPHAL-CC13}

No \hb~is detected in WESTPHAL-CC13, which may be surprising in light
of the strength of the \oiii~lines.  In MS1512-cB58, for example, the
\hb:\oiiifive~ratio is similar to the ratio between \oiii~components
at 5007\AA~and 4959\AA.  In WESTPHAL-CC13, \oiii~$\lambda 4959$~is
detected with almost 3$\sigma$~confidence at 0.33 times the strength
of the strength of \oiiifive, agreeing within the errors with the
expected fraction 1/2.87. It is unlikely that underlying stellar
absorption can account for the lack of \hb~emission.  Our
$2.5\sigma$~upper limit on the equivalent width ($EW$) is 19\AA~in the
rest frame, while stellar absorption has $EW < 5$\AA.

A likely explanation for the lack of \hb~in WESTPHAL-CC13 is sub-solar
metallicity.  A moderate range of metallicities ($\sim0.2 -
0.9~Z_{\odot}$) produce the highest ratios of $O/H$~emission lines.
With further increasing metallicity, oxygen serves as an effective
coolant, reducing the collisional excitation rates that would
otherwise increase the oxygen line strengths.  Below some critical
value ($Z\sim 0.2 Z_{\odot}$, the exact value depending on the
ionization parameter), with decreasing metallicity the absolute lack
of oxygen ions reduces the observed line emission.  A ``turn-around''
region lies near $Z\sim 0.3 Z_{\odot}$~(see KKP99).

What is the metallicity of WESTPHAL-CC13 likely to be?  We cannot
measure the oxygen abundance with the current data.  However, we can
use the non-detection of \hb~together with well established models
from the literature to place an upper limit on it.  Combining that
limit with speculation about the difference between LBGs and modern
low metallicity galaxies suggests an allowed region for the oxygen
abundance of WESTPHAL-CC13.

To properly measure
the oxygen abundance of WESTPHAL-CC13 we would need a measure of \oii~emission
as well as \oiii.  The \oiii:\oii ratio depends on the ionization
parameter, $U$~(c.f. McGaugh 1991).  For reasonable value of 
$10^{-4}<U<10^{-1}$, we expect 
$0.1<\mbox{\oiii}_{4959+5007}:\mbox{\oii} <10$~(KKP99).  Using our 
$2.5\sigma$~lower limit for \oiiifive :\hb$\sim 3.9$, we can
bound the allowed region of in the plane of metallicity vs.
emission-line ratios.  The oxygen abundance can be parameterized in
terms of the quantity 
$R_{23}\equiv [I_{3727}+I_{4959}+I_{5007}]/\mbox{H}\beta$ 
~(c.f McGaugh 1991). 
Figure \ref{fig: r23}~shows the region of allowed values for WESTPHAL-CC13.

Following KKP99, there may be an additional constraint on the
metallicity.  In nearby luminous galaxies there is a correlation
between metallicity and luminosity (Zaritsky, Kennicutt, \& Huchra,
1994).  For objects more luminous than $M_B\simeq -18$, observed
metallicities are $12+log(O/H) \gsim 8.3$.  Since LBGs share many
characteristics with moderate metallicity, low redshift starbursts, it
is likely that they follow this same correlation, but it has not yet
been measured.  Similarly, chemical evolution models of LBGs suggest a
lower limit on their metallicity of $Z \gsim 0.2 Z_{\odot}$~(Shu et al.
2000).  So the oxygen abundance of WESTPHAL-CC13 is most likely to lie
in the range of $0.25-0.8Z_{\odot}$.

In Figure \ref{fig: r23}, we also plot the metallicity measured for
MS1512-cB58 (T2000), and the limits on the metallicity of three LBGs
from P98 for which \oiii~and \hb~are measured but \oii~is lacking.  We
find that most LBGs appear likely to have less than solar metallicity,
and yet not to lie in the extremely low $Z$~regime seen in low-mass
local galaxies.  These results are in broad agreement with the results
from a larger spectroscopic sample of LBGs using NIRSPEC and ISAAC
over a wider wavelength range (Pettini et al 2000a, in preparation).

\subsection{Star Formation Rate and Dust Extinction}    
   
Balmer lines (in particular \ha, \hb) have been shown to be good
tracers of star formation in galaxies (c.f. Kennicutt 1983).  The
strength of oxygen lines is substantially more complicated, being
strongly dependent on the temperature of the ionized gas which in turn
depends on the metallicity of the galaxy.  So far it appears that many
LBGs have roughly similar metallicities, so some correlation may exist
between \oiiifive~strength and the star formation rate (SFR).  We explore
the implications with the understanding that the analysis is
inherently limited by the unknown properties of the galaxies.  

Kennicutt (1983) connected \ha~luminosity to the SFR by      
\begin{equation} 
\mbox{SFR}(\mbox{\Msun yr}^{-1}) = L(\mbox{\rm{H}}\alpha) / 1.12\times     
10^{41}\mbox{ergs~s}^{-1} .   
\end{equation}                
In order to estimate the SFR, we must assume an \oiiifive : \ha~ratio;
this assumption is uncertain due to the unknown metallicity.
MS1512-cB58 is found to have a \oiiifive : \ha~ratio of unity (and
$Z\sim 1/3~Z_{\odot}$).  We will take this value as typical for this
analysis.  Many local, strongly line-emitting, low metallicity
galaxies are seen to vary from this value a factor of order two or
less in either direction (e.g KKP99).  The variation in the \oiiifive
: \ha~ratio is also dependent on other parameters in addition to metallicity,
such as the effective temperature of the gas and the ionization
parameter (Kennicutt et al. 2000), so we are making a number of
assumptions when we fix the value of the ratio.  These complications
have led to the conventional wisdom that \oiii~is not the preferred
indicator of SFR (Kennicutt 1992).  We proceed with this caution in mind.

We have measured \oiii~in three LBGs in the present survey (we exclude
B20902-M11 and WESTPHAL-CC8 for which detections are less than
$2.5\sigma$.)  In addition to these, seven other LBGs have measured
\oiiifive~fluxes (P98; T2000; Iwamuro et al. 2000, hereafter I2000).
Inferred SFRs are listed in Table \ref{tab: all_lbg}.  Of the objects
with measured \oiiifive~fluxes, four also have measured Balmer
emission line fluxes.  Figure \ref{fig: oiii_hb}~compares the inferred
SFRs.  For three of the galaxies, the rates agree within the errors;
the fourth is less than $2\sigma$~discrepant.  This agreement provides
some evidence that trends observed in the SFRs inferred from
\oiii~will be born out in later observations of the Balmer lines.

The SFR may also be inferred from the UV continuum luminosity.
Assuming a $10^8$yr continuous star-formation model from the GISSEL96
spectral synthesis library (see Bruzual \& Charlot 1993), SFR=1
\Myr~produces $L_{1500}=8.7\times 10^{27}$ ergs
s$^{-1}$~Hz$^{-1}$~(Pettini et al. 2000b).  We can estimate
$L_{1500}$~from the measured broad-band photometry and the redshift
(given in Table \ref{tab: all_lbg}).  At $z\sim 3.5$, the
$\mathcal{R}$~filter measures the rest-frame continuum near
1500\AA~(see P98 for more discussion).  Table \ref{tab: all_lbg}~gives 
the SFR inferred from \oiiifive~and the UV continuum.

From the inferred SFRs, we can see that in most cases there is clear
evidence that SFRs are higher when inferred from \oiiifive~rather than
the UV.  A natural explanation for this difference is the presence of
dust extinction.  The UV continuum appears to underestimate the SFR by
an average factor of $\sim 3$.  This value is highly uncertain given
the number of assumptions necessary to use \oiii~as a star formation
indicator.  However, this extinction is consistent with some values
inferred from the UV continuum slopes of LBGs (e.g. P98, Steidel et
al. 1999), though it is lower than other estimates in the literature
(e.g. Trager et al. 1997, Sawicki \& Yee 1998).  The difference may be
attributable to the problems with inferring the SFR from the
\oiiifive~flux.  On the other hand, there may be a more fundamental
difference between the SFR inferred from the UV continuum and that
inferred from nebular emission lines (see Bechtold et al.  1997).

P98 observed a trend in dust extinction as a function of
(G-$\mathcal{R}$) color by comparing the ratio of SFRs inferred from
\hb~and $f_{1500}$.  We can make the same comparison using our SFRs
(see Figure \ref{fig: gr}). In order to properly utilize the
(G-$\mathcal{R}$) color intrinsic to each galaxy, we must correct the
observed color for the intervening Lyman alpha forest blanketing.  To
estimate this correction, we convolve a spectra synthesis model (from
GISSEL96, see Bruzual \& Charlot 1993) with the filter response and
the Lyman series decrements from Madau (1995).  The correction is 
quite large ($>0.5$~mag. at $z>3$).  The colors in Figure
\ref{fig: gr}~have this correction applied. In addition, we have
applied a small color term to translate the I2000 data
points from WFPC2 colors (from the Hubble Deep Field North, see
Williams et al. 1996) to (G-$\mathcal{R}$) colors (see Steidel \&
Hamilton 1993 for a description of those filters).  The color term was
determined by integrating under the filter curves using the spiral
galaxy spectrum used initially to estimate a photometric redshift for
the galaxies (see Fernandez-Soto et al. 1999, who used the bluest
spectrum from the Coleman, Wu \& Weedman 1980 library).  

We note that the trend observed in P98 from SFRs inferred using the
\hb~line in comparison to the UV is still seen when using their
\oiii~measurements as an SFR indicator.  It is difficult to judge the
trend quantitatively given the many uncertainties in the SFR estimates
(not the least of which is the systematic uncertainty in using
\oiii~as an SFR indicator).  However, taking the SFR ratios and
(G-$\mathcal{R}$) colors at face value (without error bars), we can
ask what is the probability that a random sample would have the same
appearance of a correlation.  The linear correlation coefficient (c.f.
Bevington, 1969) for the 11 data points is $r = 0.70$, which implies
that 2\%~of the time a randomly drawn population of 11 data points
would exceed this correlation.  The P98 points by themselves have a
5\%~probability of random apparent correlation, while the new points
alone have 20\%.  Excluding the two points with the highest SFR ratio
increases the probability to 22\%.  Thus, the trend seen in P98 still
appears marginally significant.

Another approach to the same problem is to look at the SFR ratio as a
function of ($\mathcal{R}$-K) color (see Figure \ref{fig: rk}).
Although the objects lie in different relative positions on this plot,
a weak trend is seen.  The linear correlation coefficient indicates
that there is a 5\%~probability of a random sample having as strong a
correlation.  Removing the point with the highest SFR ratio increases
the chance of random apparent correlation to 17\%~and removing the two
highest points increases it to 29\%.

\section{Summary}

We have presented measurements of the flux of the \oiiifive~emission
line in three LBGs with no previous optical line fluxes in the
literature.  In one of the objects (WESTPHAL-CC13), we have a strong
limit on the strength of the \hb~line, which is not seen in the spectrum.
Combining these data with three other \oiii~measurements
from recent publications (T2000, I2000), we have assembled a
list of \oiii~measurements for LBGs that more than doubles the previous
sample (P98).  This combined data set is our primary result.  

We have used the \oiii~measurements to speculate on the nature of LBGs
as a class of galaxies.  From the \oiiifive : \hb~ratio in
WESTPHAL-CC18, together with similar measurements from P98 and the
reasoning of KKP99, we consider it likely that LBGs tend to lie in the
metallicity range $0.25-0.8Z_{\odot}$.  Future measurements of
\oii~$\lambda 3727$~in LBGs will place much better constraints on this
value (Pettini et al. 2000a, in preparation) The \oiiifive~line is
also a weak tracer of star formation.  Despite uncertainties in the
calibration of the SFR as a function of \oiiifive, there is a clear
trend that SFRs inferred from the oxygen line are higher than those
inferred for the same LBGs from their rest-frame UV continuum fluxes.
This difference (a ratio of $\sim 3$) is naturally attributable to
dust extinction, and the value, while uncertain, is in general
agreement with other estimates.

These results are a first step in the direction of using optical line
diagnostics for the study of LBGs.  This new approach is now possible on larger
sample with the advent of NIRSPEC on Keck II and ISAAC on the VLT.

\acknowledgements

It is a pleasure to acknowledge the hard work and dedication of the 
NIRSPEC instrument team at UCLA:  
Maryanne Anglionto, Odvar Bendiksen, 
George Brims, Leah Buchholz, 
John Canfield, 
Kim Chim,
Jonah Hare, Fred Lacayanga,
Samuel B. Larson, Tim Liu, 
Nick Magnone,
Gunnar Skulason,
Michael Spencer, 
Jason Weiss, Woon Wong.  
In addition we thank director Fred Chaffee,
instrument specialist Thomas A. Bida, 
and the Observing Assistants
 at Keck observatory:  Joel Aycock, 
Gary Puniwai, Charles Sorenson, Ron Quick,
and Wayne Wack.

\references

\reference{} Bechtold, J., Yee, H.K.C., Elston, R., \& Ellingson, E.
1997, ApJLetters, 477, 29

\reference{} Bertin, E. \& Arnouts, S. 1996, \aaps, 117, 393

\reference{} Bevinton, P. R. 1969, {\it Data Reduction and Error
Analysis for the Physical Sciences}, McGraw-Hill Book Company, New
York, New York

\reference{} Bruzual, A. G. \& Charlot, S. 1993, \apj, 405, 538

\reference{} Bunker, A. J., Warren, S. J., Hewett, P. C., \& Clements,
D. L.  1995, MNRAS, 273, 513

\reference{} Casali, M.M. \& Hawarden, T.G. 1992, JCMT-UKIRT Newsl., No. 4, 33

\reference{} Coleman, G. D., Wu, C. C., \& Weedman, D. W. 1980, \apjs, 43, 393

\reference{} Fernandez-Soto, A., Lanzetta, K. M.,\& Yahil, A. 1999, ApJ, 513, 34

\reference{} Groth, E.J., Krisian, J.A., Lynds, R., O'Neil, E.J.,
Balsano, R., Rhodes, J., \& the WFPC-1 IDT 1994, BAAS, 26, 1403

\reference{} Howell, S.B. 1989, PASP, 101, 616

\reference{} Iwamuro, F. et al. 2000, PASJ in press (I2000;  astro-ph/0001050)

\reference{} Kennicutt, R.C. Jr. 1983, ApJ 272, 54

\reference{} Kennicutt, R.C. Jr. 1992, ApJ, 388, 310

\reference{} Kennicutt, R.C. Jr., Bresolin, F., French, H., \& Martin,
P. 2000, in press, (astro-ph/0002180)

\reference{} Kobulnicky, H.A., Kennicutt. R.C. Jr., \& Pizagno, J.L. 1999, ApJ, 514, 544 (KKP99)

\reference{} Kurucz, R. L. 1993, CD-ROM 13, ATLAS9 Stellar Atmosphere
Programs and 2 km/s Grid (Cambridge: Smithsonian Astrophys. Obs.)

\reference{} Leitherer, C., Vacca, W.D., Conti, P.S., Filippenko, A.
V., Robert, C., Sargent, W. L. W. 1996, ApJ, 465, 717

\reference{} Lowenthal, J.D. et al. 1997, ApJ, 481, 673

\reference{} Madau,P. 1995, ApJ 441, 18

\reference{} Matthews, K., \& Soifer, B. T. 1994, in {\it Infrared
  Astronomy with Arrays: The Next Generation}, ed I. McLean
(Dordrecht:Kluwer), 239

\reference{} McGaugh, S. 1991, ApJ, 380, 140

\reference{} McLean, I. S., et al. 1998, SPIE, Vol. 3354, 566

\reference{} McLean, I.S. et al. 2000, PASP submitted

\reference{} Pettini, M., Kellogg, M., Steidel, C.C., Dickinson, M.,
Adelberger, K.L., \& Giavalisco, M. 1998, ApJ, 508, 539 (P98)

\reference{} Pettini, M., et al. 2000a, in preparation

\reference{} Pettini, M., Steidel, C.C., Adelberger, K.L., Dickinson,
M., \& Giavalisco, M.  2000b, ApJ, 528, 96

\reference{} Rola, C.S., Terlevich, E., \& Terlevich, R.J. 1997, MNRAS, 289, 419

\reference{} Sawicki, M. \& Yee, H. K. C. 1998, AJ, 115, 1329

\reference{} Seitz, S., Saglia, R.P., Bender, R., Hopp, U., Belloni,
P., \& Ziegler, B.  1998, MNRAS 298, 945

\reference{} Shu, C. 2000, A\& A, in press (astro-ph/0002388)

\reference{} Steidel, C.C. \& Hamilton, D. 1993, AJ, 105, 2017

\reference{} Steidel, C.C., Giavalisco, M., Pettini, M., Dickinson,
M., \& Adelberger, K.L. 1996, ApJ Letters, 462, L17

\reference{} Steidel, C.C.,Adelberger, K.L.,Giavalisco, M.,Dickinson,
M.,\& Pettini, M.  1999, ApJ, 519, 1

\reference{} Teplitz,H.I, Malkan,M.A., \& McLean,I.S. 1998, ApJ, 506, 519 (TMM98)

\reference{} Teplitz, H.I. et al. 2000, ApJLetters, 533, 65 (T2000)

\reference{} Thompson, D., Djorgovski, S, \& Trauger, J. 1995, AJ, 110, 963

\reference{} Thompson, D.; Mannucci, F.; Beckwith, S. V. W. 1996, AJ, 112, 1794

\reference{} Trager, S.C., Faber, S.M.,Dressler, A., \& Oemler, A.
1997, ApJ, 485, 92

\reference{} Yee, H.K.C., Ellingson, E., Bechtold, R.G.,
Carlberg,R.G., Cuillandre, J.-C. 1996, AJ, 111, 1783

\reference{} Williams, R.E. et al. 1996, AJ, 112, 1335

\reference{} Zaritsky, D., Kennicutt, R.C., \& Huchra, J.P. 1994, ApJ, 420, 87

\clearpage

\begin{deluxetable}{llll}
\tablecolumns{4}

\tablecaption{Imaging Observations}
\tablehead{
\colhead{Field Target} &
\colhead{date} &
\colhead{$t_{K^{\prime}}$ (s)} &
\colhead{$t_{n.b.}$ (s)} 
}

\startdata

3C324-C6      & 1999/4/7-8 & 3240  & 11280 \\ 
B20902-M11    & 1999/4/7   & 1740  & 6480  \\
WESTPHAL-CC8  & 1999/4/7   & 1620  & 4320  \\
WESTPHAL-CC18 & 1999/4/7-8 & 3240  & 12000 \\

\enddata

\label{tab:  observations}

\end{deluxetable}

\clearpage

\begin{deluxetable}{lllllll}
\tablecolumns{7}

\tablecaption{Narrow-band Targets}
\tablehead{
\colhead{Object}&
\colhead{z}&
\colhead{K\tablenotemark{1}} &
\colhead{$SNR_K$} &
\colhead{$SNR_{n.b.}$} & 
\colhead{$\Delta m$\tablenotemark{2}} &
\colhead{EW$_{rf}$\tablenotemark{3} (\AA)}
}

\startdata

3C324-C6      & 3.310  & 22.0   & 5 & 7   & 1.6 &  $170\pm 50$ \\  
B20902-M11    & 3.300  & 20.9   & 7 & 5   & 0.2 &  $10 \pm 4$  \\ 
WESTPHAL-CC8  & 3.318  & 21.4   & 2.5 & 9 & 1.1 &  $90 \pm 50$ \\
WESTPHAL-CC18 & 3.304  & 22.2   & 6 & 6   & 1.6 &  $170\pm 50$  \\ 

\enddata
\label{tab:  nb_lbg}

\tablenotetext{1}{Magnitude on the Vega system.}
\tablenotetext{2}{The narrow-band excess, $\Delta m$, is the
  difference in magnitudes between the broad-band : narrow-band count
  ratio in the object and the ratio expected for a featureless
  continuum source (see text).}  
\tablenotetext{3}{Rest-frame equivalent width of the emission line.}
\end{deluxetable}

\clearpage

\begin{deluxetable}{ll}
\tablecolumns{2}
\tablewidth{4 in}
\tablecaption{WESTPHAL-CC13}
\tablehead{
\colhead{line}&
\colhead{flux\tablenotemark{a}}
}

\startdata

\oiiifive &  $4.6 \pm 0.9$ \nl
\oiii~$\lambda 4959$  &  $1.5 \pm 0.6$ \nl
\hb 4861    &  $< 0.47$~(1$\sigma$)

\enddata
\label{tab:  wcc13}
\tablenotetext{a}{$10^{-17}$~erg cm$^{-2}$ s$^{-1}$}

\end{deluxetable}

\clearpage

\renewcommand{\arraystretch}{.5}
\tablewidth{7.3in}
\begin{deluxetable}{lllllllll}
\tablecolumns{9}

\tablecaption{\oiiifive~flux in LBGs}
\tablehead{
\colhead{Object}&
\colhead{z} &
\colhead{$G_{AB}$}&
\colhead{$\mathcal{R}$$_{AB}$} &
\colhead{$K_{AB}$\tablenotemark{1}} &
\colhead{$f_{\lambda}$\tablenotemark{2}}&
\colhead{SFR\tablenotemark{3}}&
\colhead{SFR}&
\colhead{Ref.} \\
\colhead{} &
\colhead{} &
\colhead{} & 
\colhead{} & 
\colhead{} & 
\colhead{} & 
\colhead{(\oiii)} & 
\colhead{(1500\AA)} & 
\colhead{} 
}

\startdata

WESTPHAL-CC13                  & 3.406 & 24.7   & 23.64 & 23.1    & 4.6 $\pm$ 0.9  & 39  $^{+39 }_{-20}\pm 8$      &  32 & this work\\
3C324-C6                       & 3.310  & 25.56  & 24.73 & 23.8    & 5.4 $\pm$ 1.5  & 44  $^{+44 }_{-22}\pm 12 $      &  14 & this work\\   
B20902-M11                     & 3.300  & 25.37  & 24.19 & 22.7    & 1.25 $\pm$ .6  & 11  $^{+11 }_{-5}\pm 5$      &  22  & this work\\  
WESTPHAL-CC8                   & 3.318  & 24.69  & 24.04 & 23.2    & 4.9 $\pm$ 3    & 41  $^{+41 }_{-20}\pm 25$      &  26 & this work\\ 
WESTPHAL-CC18                  & 3.304  & 25.83  & 25.08 & 24.0    & 4.1 $\pm$ 1.3  & 34  $^{+34 }_{-17}\pm 11$      &  10 & this work\\ 
MS1512-cB58\tablenotemark{4}   & 2.739  & 24.77  & 24.29 & 23.0    & 4.9 $\pm$ 0.3  & 21  $^{+21 }_{-10}\pm 2$      &  13 & T2000\\
0000-263 D6                    & 2.971  & 23.33  & 22.88 & 22.5    & 7.6 $\pm$ 0.7  & 46  $^{+46 }_{-23}\pm 4$      &  59   & P98\\
0201+113 C6\tablenotemark{5}   & 3.053  & 24.48  & 23.90 & 23.4    & 13  $\pm$ 2.5  & 85  $^{+85 }_{-43}\pm 16$      &  23 & P98\\
0201+113 B13\tablenotemark{6}  &  2.168 & 23.37  & 23.43 & 22.9    & 6 $\pm$ 1      & 19  $\pm 3$      &  19 & P98\\
B20902-C6                      & 3.099  & 24.58  & 24.13 & 22.3    & 7.7 $\pm$ 1.1  & 53  $^{+53 }_{-26}\pm 8$      &  20 & P98\\ 
DSF2237+116C2                  & 3.333  & 24.68  & 23.55 & 22.5    & 33  $\pm$ 5    & 276 $^{+276}_{-138}\pm 42$      &  41 & P98\\
HDF 4-858.13\tablenotemark{7}  & 3.216  & 25.4   & 24.3  & 22.9    & 10.0 $\pm$ 1.1 & 77  $^{+77 }_{-38}\pm 8$      &  19 & I2000 \\
HDF 3-243.0 \tablenotemark{7}  & 3.233  & 26.7   & 25.6  & 23.8    & 5.7 $\pm$ 1.1  & 44  $^{+44 }_{-22}\pm 8$      &  6  & I2000 \\

\enddata
\label{tab:  all_lbg}

\clearpage
\tablenotetext{1}{The $K$~magnitude is given on the AB system for consistency with the optical magnitudes from
the literature.}
\tablenotetext{2}{Flux in the \oiiifive~line in units of $10^{-17}$~erg cm$^{-2}$ s$^{-1}$}
\tablenotetext{3}{Errors are the systematic and random errors respectively.  Systematic errors are the
approximately factor of two variation in the \oiiifive~: \ha~ratio.  Random errors result from photometric and
spectroscopic uncertainties.}
\tablenotetext{4}{corrected for lensing magnification (c.f. Seitz et al. 1998)} 
\tablenotetext{5}{\oiii~$\lambda 4959$~flux scaled up by the expected ratio (2.87)} 
\tablenotetext{6}{\ha~flux instead of \oiii} 
\tablenotetext{7}{F450W, F606W, F814W colors have been translated to G and $\mathcal{R}$, see text}

\end{deluxetable}
\renewcommand{\arraystretch}{1.0}

\clearpage

\figcaption[]{The narrow-band excess, $\Delta m$, vs. \kp~for objects
  in the LBG fields.  Spectroscopically confirmed LBGs are circled.
  The different symbols indicate the field in which objects were
  observed according to the key.  WESTPHAL-CC18 and 3C324 were
  observed with more integration time and so reach greater depth (see
  text).  The curved lines denote the three sigma limits above which
  an object is considered to have a narrow-band excess which indicates
  an emission line detection.  The two different lines indicate the
  separate observed depths.  The dashed line indicates the line of
  constand rest-frame equivalent width $EW_{rf}=150$\AA, and the
  dot-dash line indicates $\Delta m= 0$.
  \label{fig: cm}}

\figcaption[]{The K-band spectrum of the WESTPHAL-CC13 object,
  obtained in 2700 seconds of integration time, is shown.  Expected
  emission lines are indicated.  The dotted spectrum shows the
  1$\sigma$~errors.  The increases in the errors occur at the position
  of the night sky lines and at positions of high atmospheric
  extinction. \label{fig: kspec}}

\figcaption[]{The Strong-line ratio intensity, $log(R_{23})\equiv
  log([I_{3727}+I_{4959}+I_{5007}]/\mbox{H}\beta)$, as a function of
  oxygen abundance from KKP99, is shown.  The lines indicate the
  limiting values from reasonable ionization parameters (McGaugh et
  al. 19991).  The parameter space that the observed $R_{23}$~and
  theoretical ionization parameter constraints allow for WESTPHAL-CC13
  is shaded, with only the upper (lighter) region acceptable if the
  local observed metallicity-luminosity ratio is valid at z=3.4.  The
  data point is the value measured for MS1512-cB58 (Teplitz et al. 2000).
  The arrows indicate the limits on the metallicity of three LBGs from
  P98; for those objects no \oii~was measured, and the direction of
  the arrow shows the effect of increasing \oii~strength, from a
  minimum of \oii:\oiii$=0.1$.  \label{fig: r23}}

\figcaption[]{The SFRs inferred from the \hb~and \oiiifive~fluxes in
  four LBGs.  Object names are indicated.  Errorbars are directly
  proportional to the errors in the flux measurement.  Only random
  errors are plotted; that is, the systematic error in SFR(\oiii) is
  not plotted.  The most accurately measured object, MS1512-cB58, has
  its SFR adjusted downward by a factor of 30 to correct for lensing
  magnification.  The solid line indicates a 1:1 correspondence.
\label{fig:  oiii_hb}}

\figcaption[]{Ratio of the SFRs implied by the luminosities of the
  optical emission lines and of the UV continuum plotted against the
  observed (G-$\mathcal{R}$) color, corrected for Lyman forest
  blanketing, assuming a $10^8$~yr old continuous star formation
  model.  Open circles indicate narrow-band \oiii~detections (we
  exclude B20902-M11 and WESTPHAL-CC8 for which detections are less
  than $2.5\sigma$); the open triangle is the NIRSPEC detection of
  \oiii~in WESTPHAL-CC13; the open diamond is the NIRSPEC detection of
  \oiii~in MS1512-cB58 (T2000); filled circles are P98 spectroscopic
  detections of \oiii; the filled square is the P98 spectroscopic
  detection of \ha; the * symbols are LBGs measured in narrow-band
  imaging by I2000.
  \label{fig: gr}}

\figcaption[]{As in Figure \ref{fig: gr}, the ratio of the SFRs
  implied by the luminosities of the optical emission lines and of the
  UV continuum plotted against the observed ($\mathcal{R}$-K) color.
  Colors are converted to the AB system.  \label{fig: rk}}

\clearpage 

\begin{figure}[t]
\hskip -1in
\parbox{5in}{\epsfysize=5in \epsfbox{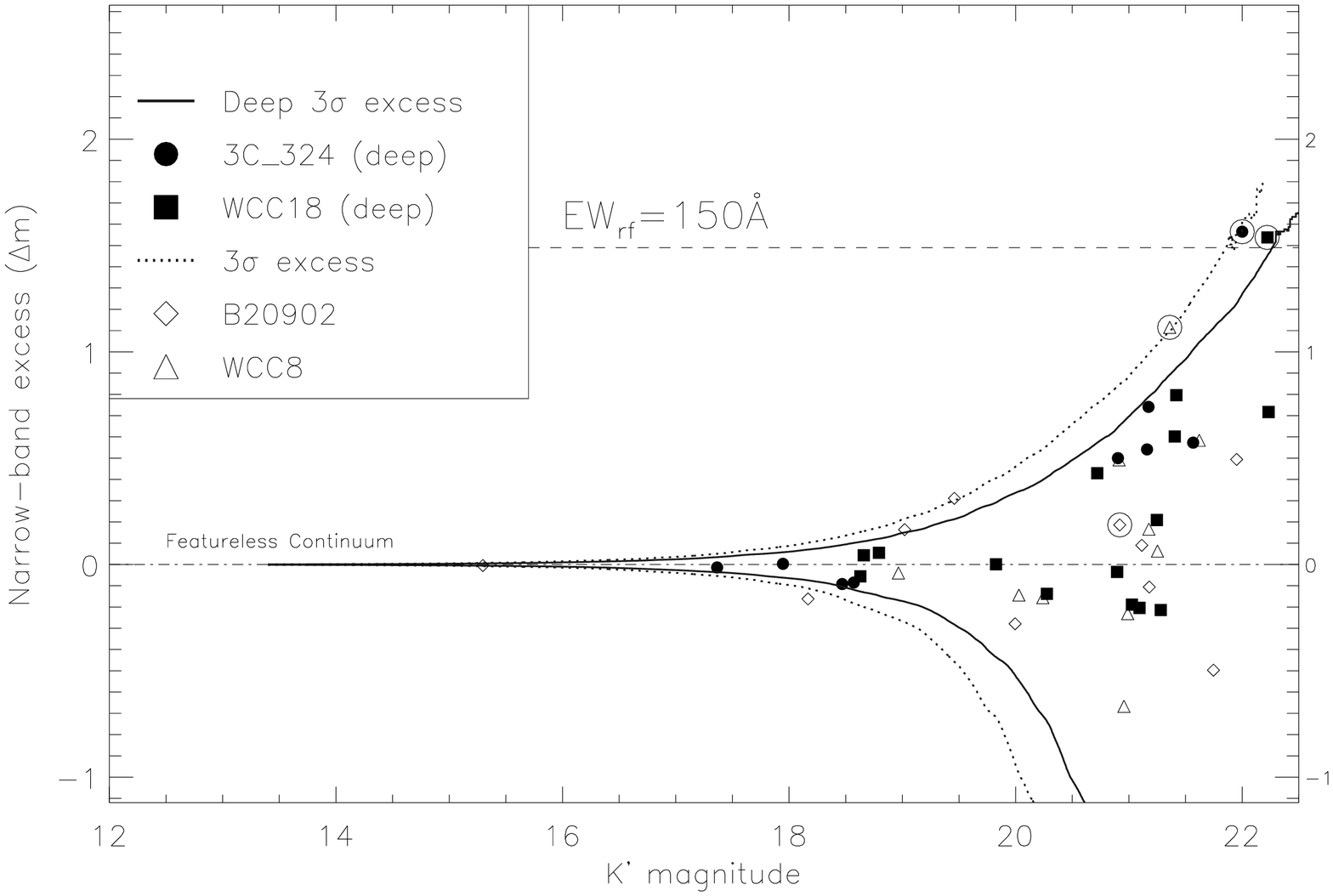}}
\end{figure}

\clearpage 

\begin{figure}[t]
\parbox{6in}{\epsfysize=6in \epsfbox{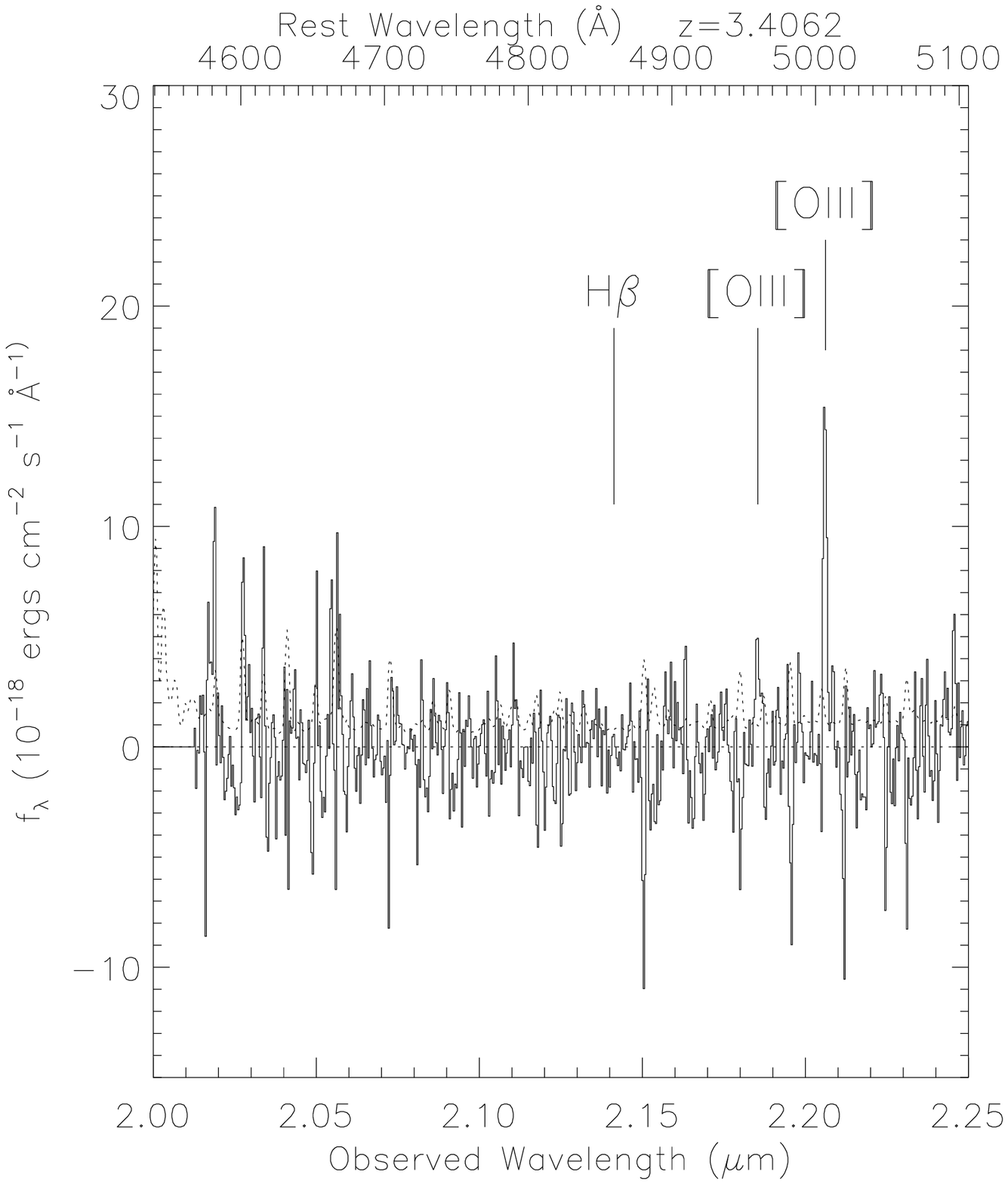}}
\end{figure}

\clearpage 

\begin{figure}[t]
\parbox{6in}{\epsfysize=6in \epsfbox{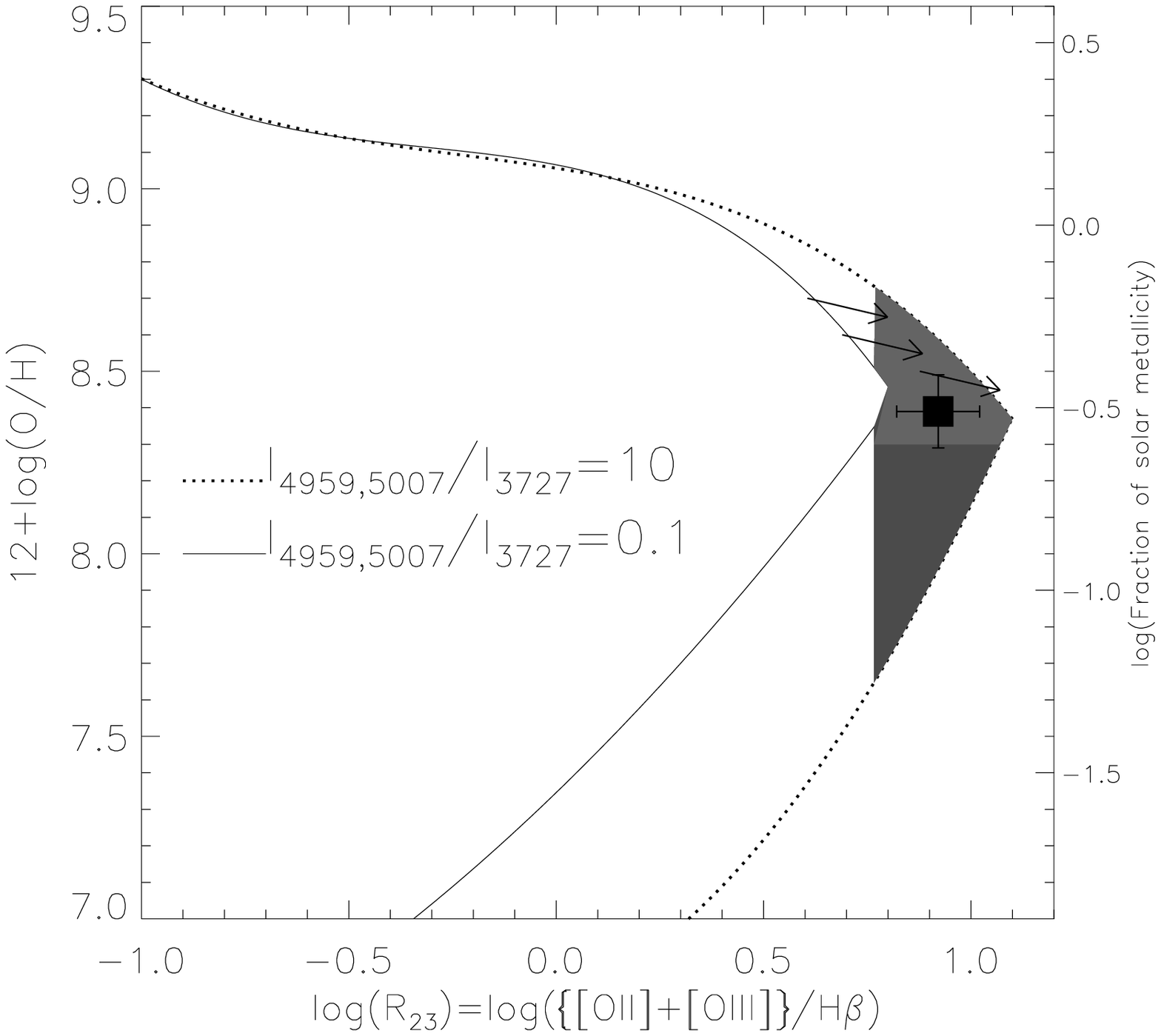}}
\end{figure}

\clearpage 

\begin{figure}[t]
\parbox{6in}{\epsfysize=6in \epsfbox{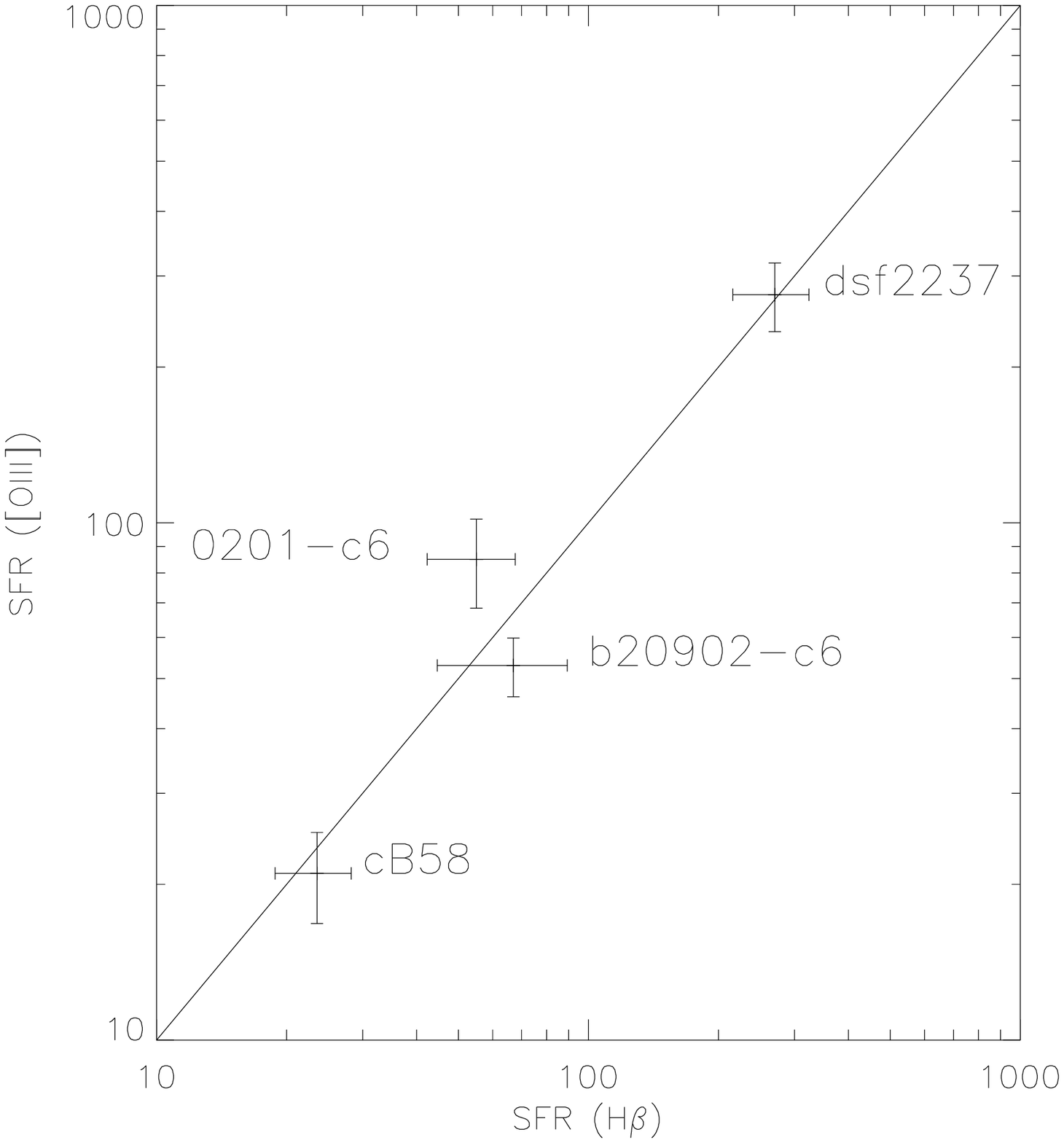}}
\end{figure}

\clearpage 

\begin{figure}[t]
\parbox{5in}{\epsfysize=5in \epsfbox{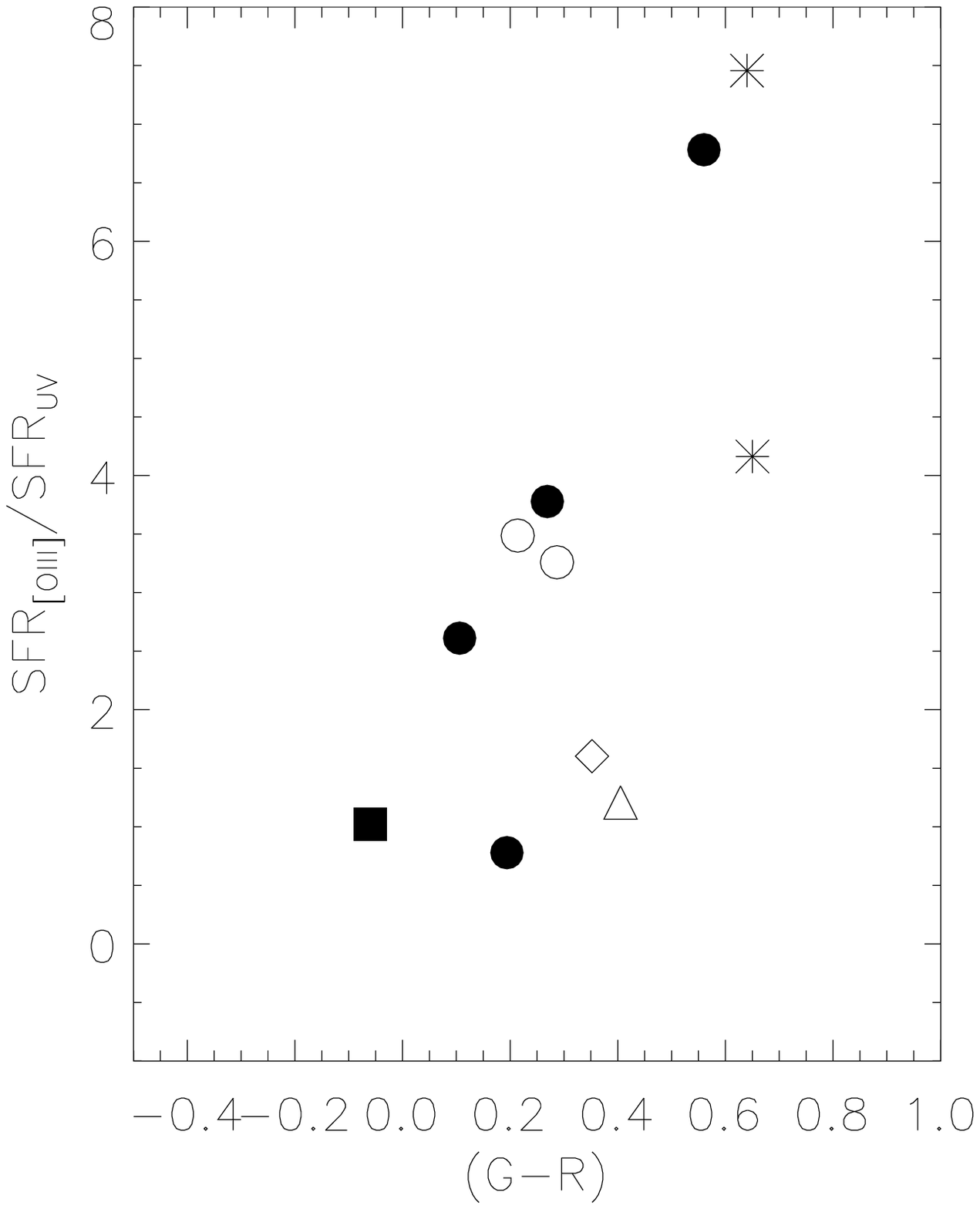}}
\end{figure}

\clearpage

\begin{figure}[t]
\parbox{5in}{\epsfysize=5in \epsfbox{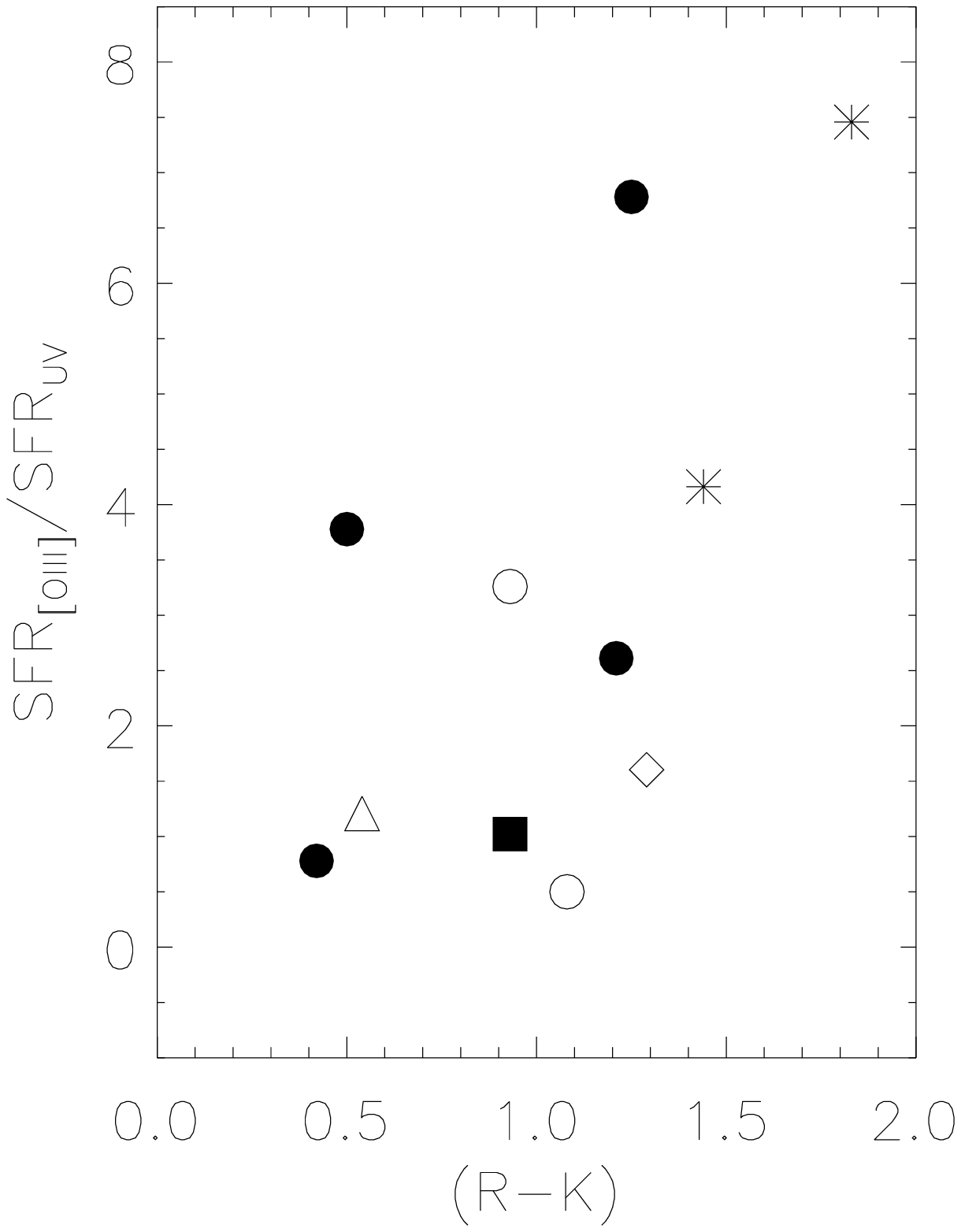}}
\end{figure}

\end{document}